\documentclass[12pt,preprint]{emulateapj}

\begin{document}

\title{Spatial Distributions of Young Stars}

\author{Adam L. Kraus\altaffilmark{1} (alk@astro.caltech.edu), Lynne A. Hillenbrand\altaffilmark{1}}
\altaffiltext{1}{California Institute of Technology, Department of 
Astrophysics, MC 105-24, Pasadena, CA 91125}

\begin{abstract}

We analyze the spatial distributions of young stars in Taurus-Auriga and 
Upper Sco as determined from the two-point correlation function (i.e. the 
mean surface density of neighbors). The corresponding power-law fits allow 
us to determine the fractal dimensions of each association's spatial 
distribution, measure the stellar velocity dispersions, and distinguish 
between the bound binary population and chance alignments of members. We 
find that the fractal dimension of Taurus is $D\sim$1.05, consistent with 
its filamentary structure. The fractal dimension of Upper Sco may be even 
shallower ($D\sim$0.7), but this fit is uncertain due to the limited area 
and possible spatially-variable incompleteness. We also find that random 
stellar motions have erased all primordial structure on scales of 
$\la$0.07$^o$ in Taurus and $\la$1.7$^o$ in Upper Sco; given ages of 
$\sim$1 Myr and $\sim$5 Myr, the corresponding internal velocity 
dispersions are $\sim$0.2 km s$^{-1}$ and $\sim$1.0 km s$^{-1}$, 
respectively. Finally, we find that binaries can be distinguished from 
chance alignments at separations of $\la$120\arcsec\, (17000 AU) in Taurus 
and $\la$75\arcsec\, (11000 AU) in Upper Sco. The binary populations in 
these associations that we previously studied, spanning separations of 
3-30\arcsec, is dominated by binary systems. However, the few lowest-mass 
pairs ($M_{prim}$$\la$0.3 $M_{\sun}$) might be chance alignments.

\end{abstract}


\section{Introduction}

The spatial distribution of young stars is a powerful diagnostic of their 
formation and early evolution. Young stars trace the gas distribution from 
which they formed, so the large-scale structure of a young association 
retains these primordial features after the gas has been accreted or 
dispersed.  On intermediate scales, the absence of structure indicates the 
typical distance over which stars have randomly dispersed since their 
birth, and therefore the velocity dispersion for the association. Finally, 
the enhanced stellar density on small scales outlines the binary 
population, distinguishing bound binary systems from chance alignments 
between young stars. Some of these topics have been addressed in previous 
work on young star distributions (Gomez et al. 1993; Larson 1995; Simon 
1997; Bate et al. 1998; Hartmann 2002), but the modern census of several 
key star-forming regions is more complete and extends to lower masses than 
a decade ago, so the analysis is worth revisiting.

The traditional tool for studying spatial distributions is the two-point 
correlation function (hereafter TPCF). The TPCF, $w(\theta)$, is defined 
as the number of excess pairs of objects with a given separation $\theta$ 
over the expected number for a random distribution (Peebles 1980). The 
TPCF is proportional to the mean surface density of neighbors, so it is 
often recast in terms of this more intuitive quantity: 
$\Sigma(\theta)=(N_*/A)[1+w(\theta)]$, where $A$ is the survey area and 
$N_*$ is the total number of stars.

In this letter, we describe an updated relation for $\Sigma(\theta)$ in 
Taurus and present the first such analysis for Upper Sco, then we fit 
power laws for the different angular regimes. Finally, we interpret our 
results to address three questions: What is the primordial fractal 
dimension of star-forming regions, and how does it relate to their 
observed geometry? What is the primordial velocity dispersion suggested by 
each association's randomization? And what is a wide binary companion, and 
can it be distinguished from an unbound association member?

\section{The Correlation Functions of Taurus and Upper Sco}

\begin{deluxetable*}{lccl}
\tablewidth{0pt}
\tablecaption{Power Law Fits}
\tablehead{
\colhead{Regime} & \colhead{Sep Range\tablenotemark{a}} & \colhead{$\alpha$} & 
\colhead{C (log deg$^{-2}$)}
}
\startdata
\multicolumn{4}{c}{Upper Sco}\\
Binary&3\arcsec-30\arcsec&-1.44$\pm$0.41&2.98$\pm$0.12 (at 9.5\arcsec)\\
Intermediate&2.8\arcmin-1.5$^o$&-0.12$\pm$0.02&1.537$\pm$0.010 (at 16\arcmin)\\
Association&1.5$^o$-4.7$^o$&-1.31$\pm$0.09&1.174$\pm$0.011 (at 2.6$^o$)\\
\multicolumn{4}{c}{Taurus}\\
Binary&3\arcsec-30\arcsec&-1.53$\pm$0.32&3.28$\pm$0.10 (at 9.5\arcsec)\\
Intermediate&1.6\arcmin-5.0\arcmin&0.12$\pm$0.39&1.62$\pm$0.05 (at 2.8\arcmin)\\
Association&5.0\arcmin-4.7$^o$&-0.951$\pm$0.007&0.650$\pm$0.005 (at 1.2$^o$)\\
\enddata
\tablenotetext{a}{There is a small range of separations between the binary 
and intermediate regimes where the data are consistent with our power law 
fits, but the uncertainties are too large for those data to contribute 
meaningfully to the fits.}
\end{deluxetable*}

\begin{figure*}
\epsscale{0.15}
\centerline{\includegraphics[width=\textwidth]{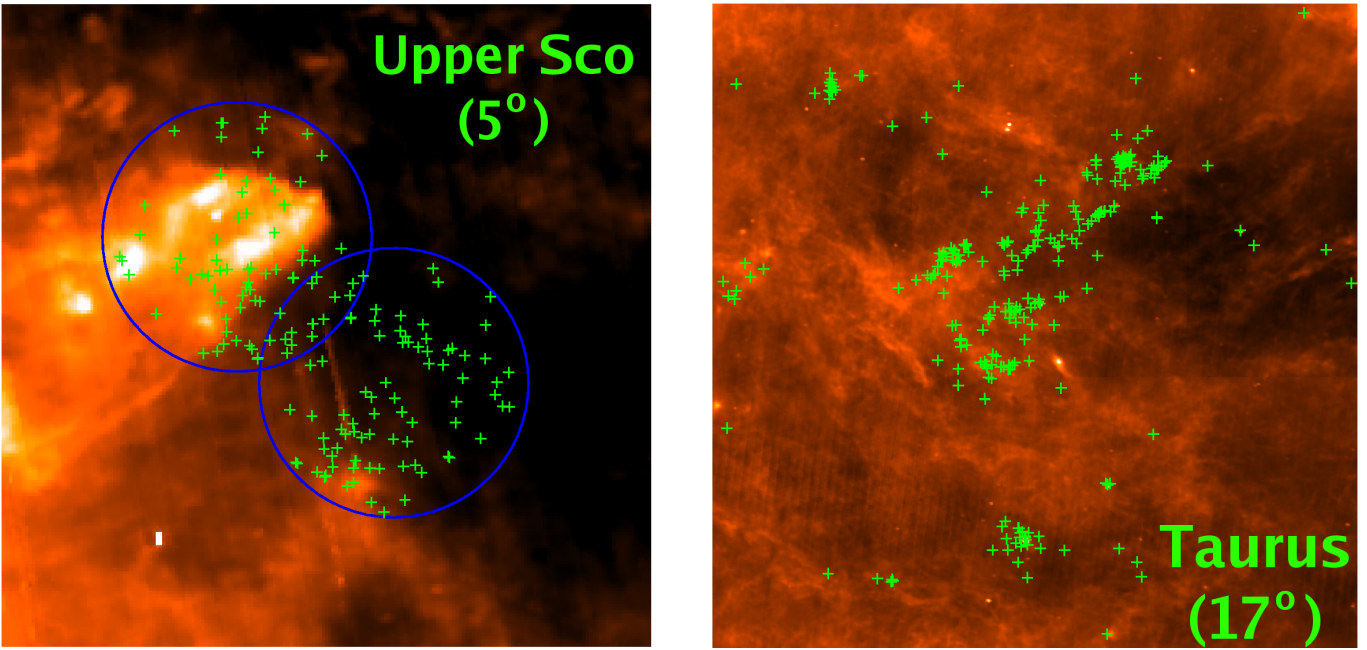}}
\caption{Locations of stars in Taurus and Upper Sco, 
superimposed on 60$\mu$m IRAS images. Members are denoted by green 
crosses, while the sample fields in Upper Sco are denoted by blue circles. 
The field of view is 17$^o$ in Taurus and 5$^o$ in Upper Sco. Known 
members in Upper Sco outline the dusty clouds in the northern field, 
suggesting systematic incompleteness for extincted members.} 
\end{figure*}

\begin{figure*}
\epsscale{1.0}
\plotone{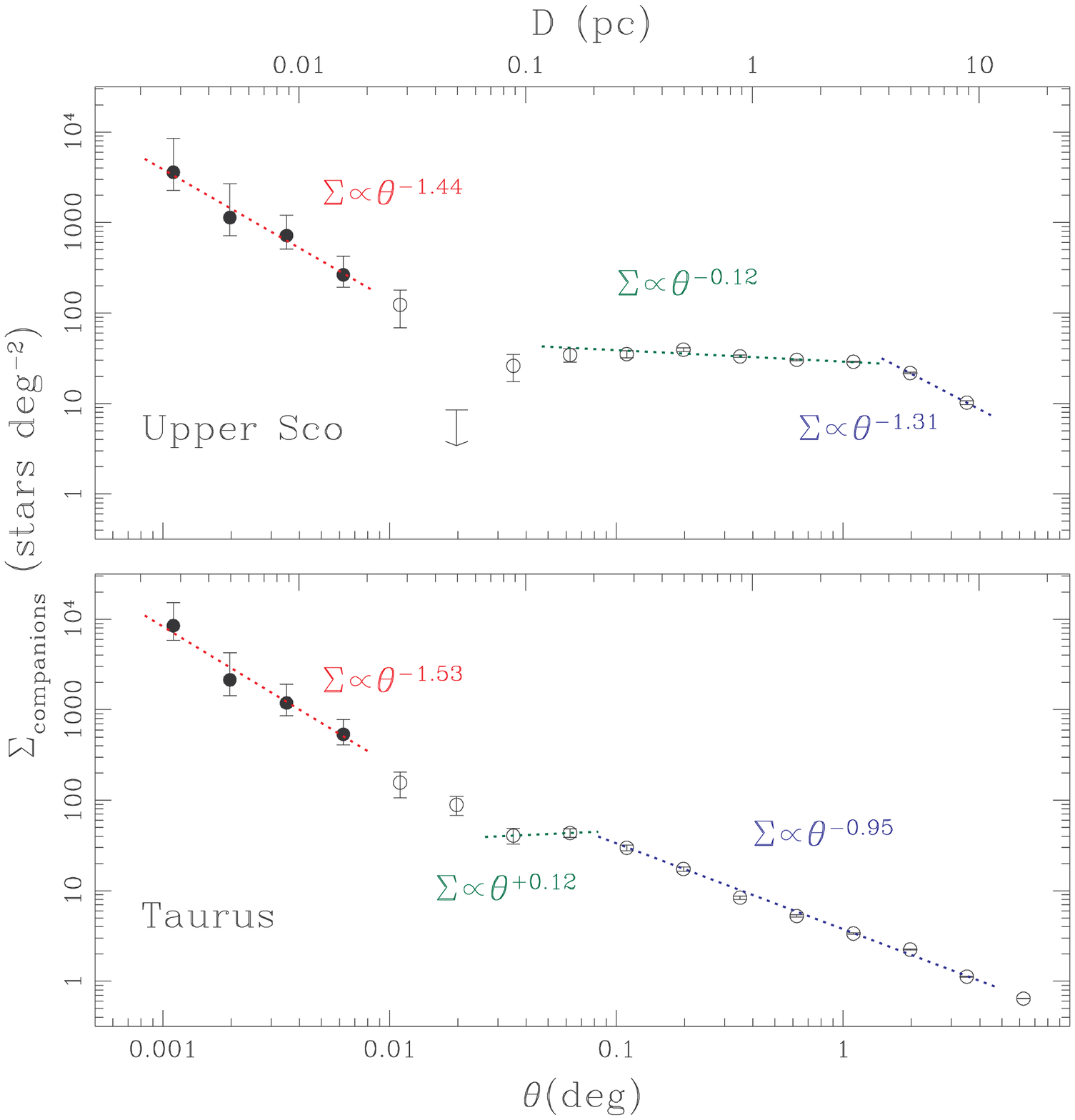}
\caption{Two-point correlation functions for members of Upper Sco and 
Taurus. These plots show the surface density of neighbors as a function of 
separation, $\Sigma$$(\theta)$, with $\theta$ in degrees (bottom axis) or 
in parsecs (top axis). The observations are from our recent wide binary 
survey (KH08; filled circles) or membership surveys in the literature 
(open circles). For each association, we have fit power laws to the 
small-scale regime (red; binary systems), the large-scale regime (blue; 
association members distributed according to the primordial structure), 
and the intermediate regime (green; association members with a randomized 
spatial distribution).}
\end{figure*}

We compiled our Taurus sample from the member list in Kraus \& Hillenbrand 
(2007a, 2008; hereafter KH07a and KH08), plus the Class 0/I sources that 
were compiled by Kenyon \& Hartmann (1995). We omitted the latter sources 
from our multiplicity surveys because their stellar properties are 
uncertain, but we include them here because that information is not 
necessary for clustering analysis. We have also included the partial list 
of new sources identified in data from the Taurus Spitzer Legacy Project 
(Padgett et al. 2006) as described by Luhman et al. (2006). For 
separations of $<$30\arcsec, we have calculated the surface density of 
neighbors only among those sources included in our initial wide binary 
survey. We have neglected the Class 0/I and heavily embedded sources 
because only some have been surveyed for multiplicity in the mid-infrared 
wavelengths (e.g. Duch\^ene et al. 2004), and not with uniform 
sensitivity. Our full sample consists of 272 members, while the 
binary-regime sample consists of 226. The Taurus sample is almost 
certainly incomplete, as a number of additional candidates have been 
identified in the Taurus Legacy Project (Padgett et al. 2006) and the XEST 
survey (Scelsi et al. 2007). However, preliminary reports suggest an 
increment of $\la$20\% in the total sample. Even if these new members do 
not trace the known distribution, their influence should be modest.

The census of Upper Sco across the full association is very incomplete, so 
we implemented our analysis for intermediate and large separations 
($\theta>30\arcsec$) using only members in two heavily-studied fields 
observed by Preibisch et al. (2002), the 2dfE and 2dfW fields. The census 
of members in these fields is not complete, but we expect that it is the 
least incomplete. As for Taurus, we calculated the surface density of 
neighbors at $<$30\arcsec\, using the full sample of our wide binary 
surveys; this choice maximizes our sample size for small separations 
(where the statistics are weakest). The 2dfE/2dfW and binary samples 
consist of 162 and 352 members.

In Figure 1, we plot the locations of our sample members superimposed on 
archival 60$\mu$m IRAS images. In Upper Sco, we see evidence of 
incompleteness for the northern field. Most of the known members outline 
the dusty regions, suggesting that any members in these regions were too 
extincted to have been identified. As we discuss later, this could affect 
the TPCF on scales of $\ga$1$^o$. In Taurus, the distribution traces the 
filamentary dust, though there are also many filaments that do not include 
any known members.

We directly measured $\Sigma(\theta)$ for Taurus because our sample spans 
the entire area of the association. However, for bounded subsets (as in 
Upper Sco), it is often easier to evaluate the TPCF via a Monte 
Carlo-based definition, $w(\theta)=N_p(\theta)/N_r(\theta)-1$, where 
$N_p(\theta)$ is the number of pairs with separations 
in a bin centered on $\theta$ and $N_r(\theta)$ is the expected number of 
pairs for a random distribution of objects over the bounded area (Hewett 
1982). The advantage is that this method does not require edge 
corrections, unlike direct measurement of $\Sigma(\theta)$. In both cases, 
we report our results as $\Sigma(\theta)$ since it is a more visually 
motivated quantity than $w(\theta)$. In Figure 2, we plot $\Sigma(\theta)$ 
for Upper Sco (top) and Taurus (bottom) spanning a separation range of 
3\arcsec\, to 10$^o$.

Based on the predicted time-evolution of young associations (Bate et al. 
1998), we expect that $\Sigma(\theta)$ can be fit with a twice-broken 
power law, corresponding to structure on three scales. At small scales, 
bound binary systems yield a relatively steep power law. At large scales 
(and for young ages, $<$1 crossing time), intra-association clustering 
yields a shallower (but nonzero) power law that corresponds to the 
primordial structure of the association. Finally, at intermediate 
separations, the random motion of association members acts to smooth out 
the primordial structure and yield a constant surface density (and thus a 
slope near zero, according to the simulations of Bate et al. 1998). The 
first knee (transition between gravitationally bound multiplicity and a 
smooth randomized distribution) corresponds to the maximum angular scale 
for distinguishing binary systems, while the second knee (transition 
between a random distribution and primordial structure) corresponds to an 
angular scale that depends on the age since members were released from 
their natal gas clouds, $\tau$, and the internal velocity dispersion, 
$v_{int}$, where $\theta$$\propto$$\tau$$v_{int}$. Hartmann (2002) 
suggested that this break also could indicate the mean spacing of cores 
along filaments (the Jeans length), which assumes that stars have 
randomized by a smaller angular scale and that the inferred value 
characteristic angular scale, the inferred value of $v_{int}$ is an upper 
limit.

In Table 1, we summarize our weighted least-squares fits for the power law 
slope $\alpha$ and zero-point $C$ in each regime. The binary regime was 
fit in the range probed in our survey of wide multiplicity (3-30\arcsec), 
while the intermediate and association regimes were fit in the ranges 
where the error bars were $\la$3\%. We established the zero point of each 
fit at the logarithmic center of the angular range in order to minimize 
correlation between $\sigma_{\alpha}$ and $\sigma_{C}$. In Upper Sco, both 
the inner and middle power laws are clearly defined, but the fit for the 
outer regime is uncertain because the angular scale is similar to the size 
of the survey area ($\sim$2-4$^o$). In Taurus, the inner and outer power 
laws are clearly defined, but the fit for the intermediate regime is 
uncertain. The TPCF at separations of 2-4\arcmin\, is flat and diverges 
from the fit for larger and smaller separations by 3-5$\sigma$, so we 
provisionally assume that this separation range represents the 
intermediate regime. The points at smaller separations also fall below the 
projection of the association-regime power law, while the points at larger 
separations agree well with the overall fit, suggesting that our inferred 
value of $v_{int}$ is at most an upper limit. The locations of the first 
knee, where the two power laws are equal, are 
$\theta_{1,USco}$$\sim$75\arcsec and $\theta_{1,Tau}$$\sim$120\arcsec; the 
respective locations of the second knee are $\theta_{2,USco}$$\sim$1.7$^o$ 
and $\theta_{2,Tau}$$\sim$0.07$^o$. The formal uncertainties in these 
measurements are only $\sim$2-3\%, but the errors are dominated by 
systematic uncertainties in the membership census and in the angular range 
over which to fit each regime.

\section{Association Regime: The Fractal Dimension of 
Taurus}

The primordial spatial distribution of young stars should trace the 
overdensities in the original gas distribution from which those stars 
formed (e.g. Hartmann 2002; Bate et al. 2003). Even if these gas 
distributions have dispersed, the remnants of primordial structure in the 
stellar distribution can still provide a key constraint to the 
distribution of overdensities during star formation. Early studies of 
TPCFs have suggested that current (and presumably primordial) stellar 
distributions are fractal in nature (e.g. Larson 1995; Simon 1997), with 
self-similar structure on a range of angular scales. Similar TPCFs can be 
reproduced (at least over a decade of separation) with simpler 
distributions like a finite number of non-fractal subclusters following a 
simple $r^\alpha$ profile (Bate et al. 1998). However, our TPCF for Taurus 
follows a single power law across $>$2 decades of separation, so it 
appears to be genuinely self-similar. The dimensionality $D$ of a fractal 
distribution indicates the extent to which it fills space, such that the 
number of neighbors $N$ within a distance $\theta$ goes as 
$N(\theta)$$\propto$$\theta$$^D$. This parameter is related to the surface 
density of neighbors; if $\Sigma(\theta)$$\propto$$\theta$$^{\alpha}$, 
then $D=\alpha$$+2$ (Larson 1995).

The fractal dimension is a result of the turbulent fragmentation that 
leads to star formation, and most models yield filamentuary structure 
(i.e. a dimension near unity). As we showed in Section 2, the observed 
power-law slope for Taurus in the large-scale regime is 
$\alpha$$=-0.951$$\pm$0.007, indicating that the fractal dimension 
($D=1.049$$\pm$0.007) is indeed close to unity. This result is consistent 
with visual inspection of the stellar distribution, as well as with CO 
maps of the remaining gas distribution (e.g. Goldsmith et al. 2008). Our 
value is significantly lower than the fractal dimension suggested by 
Larson (1995) and Simon (1997), $D=1.4$, but close to the more recent 
value suggested by Hartmann et al. (2002). Our sample is significantly 
more complete than the older samples; based on our reconstruction of those 
samples, most of the new (typically low-mass) members are located near the 
major concentrations (e.g. Strom \& Strom 1994; Brice\~no et al. 2002) 
rather than in the more distributed population (e.g. Slesnick et al. 
2006). These members increase the surface density of neighbors at small 
separations, yielding a steeper slope for $\Sigma(\theta)$. However, the 
census is still incomplete (Section 2) and if the incompleteness is 
spatially variable, such as for heavily embedded brown dwarfs, then our 
updated power-law slope could be incorrect.

We are hesitant to estimate the fractal dimension in Upper Sco. The 
appropriate regime in the TPCF includes only two separation bins, so the 
choice of bin locations could significantly affect the slope. 
Incompleteness in the dusty northern region could also influence the 
inferred large-scale structure. However, if we adopt our power-law fit 
from Section 2 ($\alpha$$=-1.31$$\pm$0.09), we find that $D=0.69\pm$0.09 
on scales of $\sim$2$^o$.

\section{Intermediate Regime: The Primordial Velocity Dispersion}

The angular scales over which structure has been randomized, as indicated 
by the location of the second knee in $\Sigma(\theta)$, directly 
constrains the primordial velocity dispersion for each association (e.g. 
Bate et al. 1998). This constraint is particularly important for 
low-density associations like Taurus and Upper Sco because the expected 
velocity dispersion ($\la$1-2 km s$^{-1}$; Frink et al. 1997) may be too 
low to be measured easily via a direct method (like high-resolution 
spectroscopy to determine radial velocities). OB and T associations are 
not bound once their unaccreted gas is expelled (e.g. Lada et al. 1984), 
so the internal velocity dispersion is critical for determining how long 
they can persist as recognizable moving groups (like the $\beta$ Pic, TW 
Hya, or $\mu$ Oph associations; Webb et al. 1999; Zuckerman et al. 2004; 
Mamajek 2006) and how long substructure can remain in these moving groups.

Allowing for projection effects, the angular scales of each TPCF's outer 
knee correspond to physical dispersion scales of $\sim$0.23 pc in Taurus 
and $\sim$6 pc in Upper Sco. Given the characteristic ages of each 
association ($\sim$1 Myr and $\sim$5 Myr, respectively), the corresponding 
characteristic velocity dispersions are $\sim$0.2 km s$^{-1}$ and 
$\sim$1.0 km s$^{-1}$. As we previously discussed, there is uncertainty in 
the fits, so this values should be taken with caution. We also note that 
these values represent the velocity dispersion with respect to other stars 
only within an angular distance of $\sim$$\theta_{knee}$. We can't rule 
out the possibility that larger substructures are moving coherently with a 
higher velocity dispersion, only that any substructure with angular size 
$\theta$ is not moving with sufficient speed 
($\dot{\theta}$$\sim$$\theta$$/\tau$) that its angular displacement from 
birth is of order $\theta$. This limit also suggests an explanation for 
the larger velocity dispersion in Upper Sco; even if the velocity 
dispersion within $\sim$0.1-0.2 pc substructures is the same as in Taurus, 
the observed TPCF could be reproduced if the velocity dispersion between 
those substructures is $\sim$1 km s$^{-1}$. A scale dependence in the 
velocity dispersion could also explain previous proper motion studies in 
Taurus, which found velocity dispersions within the major subclumps (on 
scales of $\sim$1-3 pc; e.g. Jones \& Herbig 1979) that were $\sim$1 km 
s$^{-1}$.

A similar effect has been noted in locations like the ONC, where radial 
velocities show an overall north-south gradient of $\sim$5 km s$^{-1}$ in 
addition to the local velocity dispersion of 2-3 km s$^{-1}$ (F\~ur\'esz 
et al. 2008). However, there is also observational evidence that 
small-scale velocity dispersions are higher in denser clusters; 
submillimeter observations of IRS1 in NGC 2264 (Williams \& Garland 2002) 
found that six protostellar cores (spanning 0.44 pc) had a velocity 
dispersion of 0.9 km s$^{-1}$, which much higher than the velocity 
dispersion that we find in Taurus, though also closer to the value for 
scales of 1-3 pc suggested by Jones \& Herbig (1979).

Our results suggest that regions like Taurus and Upper Sco are even less 
dynamically active, relative to the ONC, than their lower densities might 
imply. The velocity dispersions also provide a direct estimate of the 
virial velocity in the natal environment (before the removal of gas) and 
therefore jointly constrain the typical mass and size of a star-forming 
clump: $Mv^2$$\sim$$(3GM^2)/(5R)$ or $M/R$$\sim$$(5v^2)/(3G)$, yielding 
$M/R\sim$15 in Taurus and $M/R\sim$550 in Upper Sco, where the mass is in 
solar masses and the radius is in parsecs. Thus, the primordial 
star-forming structures that are now dispersing with these characteristic 
velocities were smaller and/or more massive in Upper Sco than in Taurus.

\section{Binary Regime: What is a Binary System?}

The existence and properties of wide binary systems are critical for 
constraining multiple star formation in the limiting case of large 
separations and early times. If wide binaries form out of a single 
protostellar clump, then the maximum separation also constrains the 
maximum size of clumps that can collapse to become bound systems. As 
previous authors have suggested (e.g. Larson 1995), the outer edge of the 
young binary separation distribution is similar to the mean Jeans length 
for nearby molecular clouds. This limit is also similar to the maximum 
separation seen in the field (e.g. Duquennoy \& Mayor 1991), suggesting 
that some wide binaries join the field without being subjected to 
significant dynamical interactions. However, study of young binaries is 
complicated by the difficulty of distinguishing gravitationally bound 
binary pairs from coeval, comoving association members that are aligned in 
projection. We addressed this issue for a single system in Upper Sco 
(UScoJ1606-1935; Kraus \& Hillenbrand 2007b) by calculating the 
association's TPCF to determine the probability that it is a bound system; 
we now extend our analysis to the full known populations of Taurus and 
Upper Sco.

We find that the transition between the binary and intermediate regimes 
occurs at $\sim$11000 AU in Upper Sco and $\sim$17000 AU in Taurus. The 
binary population therefore extends at least to these angular scales, but 
we can not distinguish binary companions from chance alignments outside 
this limit. The difference between these regimes is a result of the higher 
total wide binary frequency in Taurus (KH07a, KH08), as the overall 
surface density of ``contaminant'' co-association members is similar in 
both associations. The number statistics do not support any assertions 
regarding the outer maximum limit of binary formation, but this angular 
scale matches both the maximum binary separation seen in the 
field and the typical Jeans length, so we do not expect to find many 
binary systems with wider separations.

Candidate companions inside this limit could also be chance alignments, 
but the probability drops for progressively smaller separations. In Upper 
Sco, we expect $\sim$3.3 chance alignments with separations of 
15-30\arcsec\, from an intermediate- or high-mass member ($M_{prim}>$0.4 
$M_{\sun}$), plus another $\sim$2.4 chance alignments of two low-mass 
members. The number of high-mass chance alignments is far lower than the 
total number of pairs, which suggests that the vast majority are bound 
binaries. However, our wide binary survey found only four wide pairs of 
low-mass companions, so it is unclear whether any are genuine binary 
systems. The contamination rate is moderately lower in Taurus, yielding 
$\la$1 contaminant in either mass range, but the results are similar. Most 
of the high-mass pairs are binary systems, but the two low-mass pairs may 
or may not be bound binary systems.

\acknowledgements

ALK is supported by a NASA Origins grant to LAH.


\begin{thebibliography}{}
\bibitem[Bate et al.(1998)]{bate98} Bate, M., Clarke, C., \& McCaughrean, M. 1998, \mnras, 297, 1163
\bibitem[Brice\~no et al.(2002)]{bric02} Brice\~no, C., Luhman, K., 
Hartmann, L., Stauffer, J., \& Kirkpatrick, J. 2002, \apj, 580, 317
\bibitem[Duch\^ene et al.(2004)]{duc04} Duch\^ene, G., Bouvier, J.,
Bontemps, S., Andr\'e, P., \& Motte, F. 2004, \aap, 427, 651
\bibitem[Duquennoy \& Mayor(1991)]{dm91} Duquennoy, A. \& Mayor, M. 1991, \aap, 248, 485
\bibitem[Frink et al. (1997)]{frink97} Frink, S., Roser, S, Neuhauser, R., \& Sterzik, M. 1997, 
\aap, 325, 613
\bibitem[F\~ur\'esz et al.(2008)]{fur08} F\~ur\'esz, G., Hartmann, L., 
Megeath, S., Szentgyorgyi, A., \& Hamden, E. 2008, \apj, 676, 1109
\bibitem[Goldsmith et al.(2008)]{g08} Goldsmith, P., Heyer, M., 
Narayanan, G., Snell, R., Li, D., \& Brunt, C. 2008, arXiv:0802.2206
\bibitem[Gomez et al.(1993)]{gom93} Gomez, M., Hartmann, L., Kenyon, S., 
\& Hewett, R. 1993, \aj, 105, 1927
\bibitem[Hewett(1981)]{hew81} Hewett, P. 1982, \mnras, 201, 867
\bibitem[Jones \& Herbig(1979)]{jh79} Jones, B. \& Herbig, G. 1979, \aj, 
84, 1872
\bibitem[Kraus \& Hillenbrand(2007a)]{kraus07a} Kraus, A. \& Hillenbrand, L. 2007a, 
\apj, 662, 413 (KH07a)
\bibitem[Kraus \& Hillenbrand(2007b)]{kraus07b} Kraus, A. \& Hillenbrand, L. 2007b, 
\apj, 664, 1167
\bibitem[Kraus \& Hillenbrand(2008a)]{kraus08a} Kraus, A. \& Hillenbrand, L. 2008, 
\apj, submitted (KH08)
\bibitem[Lada et al.(1984)]{lada84} Lada, C., Margulis, M., Dearborn, D. 
1984, \apj, 285, 141
\bibitem[Larson(1995)]{lar95} Larson, R. 1995, \mnras, 272, 213
\bibitem[Luhman et al.(2006)]{luh06b} Luhman, K., Whitney, B., Meade, M., 
Babler, B., Indebetouw, R., Bracker, S., \& Churchwell, E. 2006, \apj, 647, 1180
\bibitem[Mamajek(2006)]{mam06} Mamajek, E. 2006, \aj, 132, 2198
\bibitem[Padgett et al.(2006)]{pad06} Padgett, D. et al. 2006, BAAS, 209, 3016
\bibitem[Peebles(1980)]{peeb80} Peebles, J. 1980, The Large Scale Structure of 
the Universe (Princeton: Princeton Univ. Press)
\bibitem[Preibisch et al.(2002)]{pre02} Preibisch, T. et al. 2002, \aj, 124, 404
\bibitem[Scelsi et al.(2007)]{scel07} Scelsi, L. et al. 2007, \aap, 468, 
405
\bibitem[Simon et al.(1996)]{sim96} Simon, M., Holfeltz, S., Taff, L. 1996, 
\apj, 469, 890
\bibitem[Simon(1997)]{sim97} Simon, M. 1997, \apj, 482, 81
\bibitem[Slesnick et al.(2006)]{sles06b} Slesnick, C., Carpenter, J.,
Hillenbrand, L., \& Mamajek, E. 2006, \aj, 132, 2665
\bibitem[Strom \& Strom(1994)]{ss94} Strom, K. \& Strom, S. 1994, \apj, 424, 237
\bibitem[Webb et al.(1999)]{webb99} Webb, R., Zuckerman, B., Platais, I., 
Patience, J., White, R., Schwartz, M., \& McCarthy, C. 1999, \apj, 512, 63
\bibitem[Williams \& Garland(2002)]{wg02} Williams, J. \& Garland, C. 2002, 
\apj, 568, 259
\end{thebibliography}
\end{document}